\documentclass[useAMS,usenatbib,aas_macros]{mn2e}
\usepackage{aas_macros}
\usepackage{amsmath}

\usepackage{color}
\newcommand{\adb}[1]{{#1}}

\newcommand{\equ}[1]{eq.~(\ref{eq:#1})}
\newcommand{\equs}[1]{eqs.~(\ref{eq:#1})}

\newcommand{\se}[1]{\S\ref{sec:#1}}

\newcommand{\Fig}[1]{Figure~\ref{fig:#1}}

\newcommand{\be}{\begin{equation}}
\newcommand{\ee}{\end{equation}}
\newcommand{\bea}{\begin{eqnarray}}
\newcommand{\eea}{\end{eqnarray}}
\def\ra{\rangle}
\def\la{\langle}

\newcommand{\msun}{M_\odot}

\newcommand{\ifm}[1]{\relax\ifmmode#1\else$\mathsurround=0pt #1$\fi}
\newcommand{\kms}{\ifmmode\,{\rm km}\,{\rm s}^{-1}\else km$\,$s$^{-1}$\fi}

\newcommand{\kpc}{\,{\rm kpc}}

\newcommand{\Gyr}{\,{\rm Gyr}}

\newcommand{\Myr}{\,{\rm Myr}}

\newcommand{\ltsima}{$\; \buildrel < \over \sim \;$}
\newcommand{\lsim}{\lower.5ex\hbox{\ltsima}}
\newcommand{\gtsima}{$\; \buildrel > \over \sim \;$}
\newcommand{\gsim}{\lower.5ex\hbox{\gtsima}}
\newcommand{\prop}{\propto}

\def\Mv{M_{\rm v}}

\def\Rv{R_{\rm v}}
\def\Vv{V_{\rm v}}

\def\Mg{M_{\rm g}}
\def\Ms{M_{\rm star}}

\def\Mt{M_{\rm tot}}

\def\Md{M_{\rm d}}
\def\Rd{R_{\rm d}}

\def\fg{f_{\rm g}}
\def\fc{f_{\rm c}}
\def\fs{f_{\rm s}}
\def\Mcold{M_{\rm cold}}
\def\delw{\delta_{w=1}}
\def\lamw{\lambda_{w=1}}
\def\dellam{\delta_{\lambda}}

\def\Mclump{M_{\rm clump}}

\def\Rclump{R_{\rm clump}}

\def\td{t_{\rm d}}
\def\tff{t_{\rm ff}}

\def\tinf{t_{\rm inf}}
\def\tsfr{t_{\rm sfr}}
\def\epssfr{\epsilon}

\def\Re{R_{\rm d}}

\def\fb{f_{\rm b}}
\def\md{m_{\rm d}}
\def\Sigc{\Sigma_{\rm cold}}
\def\Sigg{\Sigma_{\rm g}}
\def\Sigh{\Sigma_{\rm hot}}

\def\Sigs{\Sigma_{\rm s}}
\def\Sigw{\Sigma_{\rm s,w=1}}
\def\Sig1{\Sigma_{1\kpc}}

\def\M12{M_{\rm v,12}}
\def\R100{R_{\rm v,100}}
\def\z3{(1+z)_3}
\def\fcc{f_{\rm c,0.5}}
\def\fss{f_{\rm s,0.5}}
\def\mdd{m_{\rm d,0.05}}
\def\lamm{\lambda_{0.05}}

\def\ga{\tau_0^{-1}}

\def\rf{\par\smallskip\noindent\hangindent 10pt}

\def\bul{$\bullet\ $}

\title[Blue Nuggets by VDI]
{Wet Disc Contraction to Galactic Blue Nuggets\\
 and Quenching to Red Nuggets}

\author[Dekel et al.]
{
A. Dekel$^{1}$,
A. Burkert$^{2}$
\\
\\
$^1$Center for Astrophysics and Planetary Science,
Racah Institute of Physics, The Hebrew University, Jerusalem 91904 Israel\\
(avishai.dekel@mail.huji.ac.il)\\
$^2$Universitaets-Sternwarte Scheinerstr. 1,
Munchen, D-81679, Germany\\
and Max-Planck-Institut fuer Extraterrestrischen Physik, P.O. Box 1312, 
D-81679 Muenchen, Germany, Max-Planck-Fellow\\
(burkert@usm.lmu.de)
}
\begin{document}

\large

\pagerange{\pageref{firstpage}--\pageref{lastpage}} \pubyear{2002}

\maketitle

\label{firstpage}

\begin{abstract}
\adb{
We study the origin of high-redshift, compact, quenched spheroids (red nuggets)
through the dissipative shrinkage of gaseous discs into compact star-forming
systems (blue nuggets).  The discs, fed by cold streams, undergo violent disc
instability (VDI) that drives gas into the centre (along with mergers). The
inflow is dissipative when its timescale is shorter than the star formation
timescale. This implies a threshold of $\sim\!0.28$ in the cold-to-total mass
ratio within the disc radius.  For the typical gas fraction
$\sim\!0.5$ at $z\!\sim\!2$, this threshold is traced back to a maximum spin
parameter of $\sim\!0.05$, implying that $\sim$half the star-forming galaxies
contract to blue nuggets, while the rest form extended stellar discs.
Thus, the surface density of blue galaxies is expected to be bimodal about
$\sim\!10^{9}\msun\kpc^{-2}$, slightly increasing with mass.
Blue nuggets are expected to be rare at low $z$ when the gas fraction is low.
The blue nuggets quench to red nuggets by complementary internal and external
mechanisms.  Internal quenching by a compact bulge, in a fast mode and
especially at high $z$, may involve starbursts, stellar and AGN feedback,
or Q-quenching.  Quenching due to hot-medium haloes above $10^{12}\msun$
provides maintenance and a slower mode at low redshift.  These predictions
are confirmed in simulations and are consistent with observations at
$z\!=\!0\!-\!3$.
}
\end{abstract}


\section{Introduction}
\label{sec:intro}

Observations indicate that a significant fraction of the massive galaxies
at redshifts $z=2-3$ are compact ellipticals with suppressed star formation
rates (SFR), for which we adopt the nickname ``red nuggets"  
\citep{dokkum08,damjanov09,dokkum09,newman10,damjanov11,whitaker12}. 
While the massive star-forming discs of stellar mass $\sim 10^{11}\msun$ 
extend to effective radii of several kpc
\citep{genzel06,genzel08},
the quenched spheroids of a similar mass have effective radii of order 1 kpc.
\adb{
The disc sizes are roughly consistent with the straightforward theoretical 
expectations based on gas infall through the dark-matter haloes into
rotating discs while conserving angular-momentum \citep{mmw98,bullock01_j}.
Since stellar systems tend to conserve energy and angular momentum,
further contraction to form the compact nuggets would require
further loss of energy and loss of angular momentum, which cannot be easily
achieved by stellar systems. Thus, the formation of nuggets is likely to be
a dissipative process, 
} 
namely associated with gas inflow into the central regions of the galaxies.
Gas inflow is expected to be generated naturally at high redshift, 
where the gas fraction in
the discs is high to begin with \citep{daddi10,tacconi10,tacconi13}.
Indeed, there are indicative observational identifications of the progenitors 
of the red nuggets in the form of ``blue nuggets", which are compact, 
star-forming galaxies with consistent abundances 
\citep[][Bruce et al. in preparation]{barro13,williams13,barro13b},
\adb{
though the observed abundances of blue nuggets depend on their lifetimes, 
which could be rather short, depending on the galaxy properties and on the 
actual quenching mechanism (see below).
}

\smallskip
The chain of events envisioned at high redshift consists of  
(1) accretion-driven violent disc instability,
(2) dissipative contraction into compact, star-forming blue nuggets,
and
(3) quenching of star formation into compact quenched red nuggets. 
This may be followed by (4) a gradual expansion of the ellipticals.

\smallskip
In phase (1), streams from the cosmic web, consisting of smooth gas and 
merging galaxies, continuously feed the discs
\citep{bd03,keres05,db06,ocvirk08,keres09,dekel09}.
\adb{
The detailed thermal history of the streams in the inner halo
\citep{cdb10,nelson13}
does not make a difference, as long as the discs are fed with cold gas
at the levels consistent with the observed high SFR and gas fraction. 
}
The high gas fraction induces VDI, which is characterized by 
turbulence and perturbations in the form of large transient features and 
giant clumps
\citep{noguchi99,immeli04_a,immeli04_b,genzel06,bournaud07c,genzel08,
dsc09,agertz09,cdb10,ceverino12}. 

\smallskip
In phase (2), dissipative gas inflows into the disc centres are naturally
driven by the VDI
\citep{gammie01,dsc09,krum_burkert10,burkert10,bournaud11,
forbes12,cdg12,elmegreen12,forbes13,dekel13}.
Similar gas inflows are generated by wet mergers
\citep[e.g.][]{barnes91,mihos96,hopkins06}.
The gas inflows within the discs produce massive compact 
gas-rich central systems that form stars at a high rate --- the blue nuggets. 
The evolution along the blue sequence
of star-forming galaxies is accompanied by gradual stellar and halo mass growth.

\smallskip
The quenching to red nuggets in phase (3) can be driven by two complementary
types of mechanisms, bulge quenching and halo quenching. 
Bulge quenching may involve, for example, starbursts, 
stellar feedback \citep[e.g.][]{ds86,murray05}, 
AGN feedback \citep[e.g.][for a review]{ciotti07,cattaneo09}, 
or morphological quenching \citep{martig09}. 
These processes typically operate in a fast mode that may be the 
dominant trigger for quenching at high redshift.
The massive halo quenching may involve, for example, virial shock heating 
\citep{bd03,keres05,db06,keres09},
gravitational infall heating
\citep{db08,khochfar08},
or AGN feedback coupled to the hot halo
\citep{db06,cattaneo09,fabian12}. 
These processes typically operate in a slow, maintenance mode that is 
dominant at low redshift or at the late stages of the quenching.

\smallskip
Finally, in phase (4), the red nuggets gradually grow in mass and 
expand by dry mergers along the red sequence. 

\smallskip
Here we focus on phase (2), the formation of blue nuggets by VDI-driven 
wet inflow.
Gas inflow that is triggered by wet mergers has been studied before
\citep[][and references therein]{hopkins06}.
However, it is becoming evident that at high redshift more gas is driven 
into the bulge by VDI inflow than by mergers, both based on observations 
\citep[e.g.][]{genzel06,genzel08,bournaud08,kaviraj13b} 
and on theory including simulations 
\citep[e.g.][]{dekel09,bournaud09,cattaneo13,dekel13}.

\smallskip
A key idea is that for the inflow within the disc to be intense and 
dissipative, the characteristic timescale for star formation, $\tsfr$, 
should be longer 
than the timescale for inflow, $\tinf$. Otherwise, most of the disc mass will 
turn into stars before it reaches the bulge, the inflow rate will be 
suppressed, and the system will become an extended stellar disc.
We thus define a ``wetness" parameter 
\be
w \equiv \frac{\tsfr}{\tinf} \, ,
\label{eq:w}
\ee
such that the condition for a wet inflow is $w>1$.

\smallskip
Once $w>1$, we will argue that
the contraction continues till the system becomes dispersion
dominated and the VDI phases out.
A general consequence is that the star-forming galaxies are expected to
show a bimodality in central density, separating the extended discs
from the compact spheroids.

\smallskip
We will also discuss the predicted distinction between fast and slow 
modes of contraction and quenching, which originate from VDI discs with
$w \gg 1$ and $w \gsim 1$ respectively.



\smallskip
At a given mass and redshift, the distribution of $w$ values,
and the subsequent variation in evolution path, 
may be dominated by variations in contraction factor from the virial
radius to the disc radius, which can be traced back to variations in
spin parameter of the baryons that make the disc. 
The threshold $w=1$ can be translated to a critical value of spin parameter,
$\lambda_{w=1}$. The fraction of star-forming galaxies that will become blue
nuggets then depends on the value of $\lambda_{w=1}$ with respect to the
average value of the spin distribution, $\la \lambda \ra \sim 0.05$, 
and it will be computed.
 
\smallskip
A strong redshift dependence of the blue-nugget fraction is expected due to
the variation of gas fraction with redshift.
The critical value of central surface density will be estimated as a function 
of redshift and mass based on scaling relations for gas fraction and 
stellar-mass fraction in haloes as estimated from observations.

\smallskip
In \se{wet_inflow} we derive the threshold condition for wet inflow
in terms of the properties that characterize the VDI phase.
In \se{bn_spin} we relate this threshold to the spin parameter and its
distribution, and estimate the redshift and mass dependence of the fraction
of blue nuggets and the surface-density threshold.
In \se{quenching} we address the possible quenching mechanisms from blue
nuggets to red nuggets.
In \se{conc} we summarize our conclusions and discuss them.

\section{VDI-driven wet inflow}
\label{sec:wet_inflow}

\subsection{Violent Disk Instability}
\label{sec:VDI}

\adb{
We summarize here the basic features of Toomre instability in a cosmological
context following the analysis and notation of \citet{dsc09}.  
}
Galactic discs at high redshift are expected to develop a 
gravitational disc instability with a Toomre instability
parameter $Q$ smaller than unity \citep{toomre64}.
The Toomre parameter can be expressed as 
\be
Q \simeq \frac{\sqrt{2}\Omega\sigma}{\pi G \Sigma} 
\simeq \frac{\sqrt{2}}{\delta} \frac{\sigma}{V} \, .
\label{eq:Q}
\ee 
The radial velocity dispersion $\sigma$ and the surface density $\Sigma$
refer to the cold component of the disc that participates in the disc 
instability. We sometimes refer to it as ``gas", but it also includes the 
``cold" young stars.
The velocity dispersion $\sigma$ represents supersonic turbulence that provides
the pressure while thermal pressure is negligible.
The angular velocity $\Omega$ and circular velocity $V$
refer to the characteristic disc radius $\Rd$, $V=\Omega \Rd$.
The constant $\sqrt{2}$ refers to a flat rotation curve 
(it stands for $\sqrt{2(1+\nu)}$ where the rotation curve is
$V(r) \prop r^\nu$).
The cold mass fraction $\delta$ is the ratio of mass in the cold component 
to the total mass within $\Rd$, including gas, young and old stars, 
and dark matter,
\be
\delta \equiv \frac{\Mcold}{\Mt} \, .
\ee
This is the key physical parameter governing the instability.

\smallskip
The instability in high-redshift galaxies
is driven by the high cold surface density that reflects the 
high mean cosmological density, the high gas accretion rate, and the
inability of the star formation rate (SFR) to catch up with the accretion
rate.
We term this phase {\it violent\,} disc instability 
\adb{
(VDI)
}
because the associated
dynamical processes occur on a galactic orbital timescale, as opposed to the
secular evolution at low redshift (see below).
The unstable disc tends to self-regulate itself in marginal instability
with $Q \simeq 1$.\footnote{For a thick disk $Q \simeq 0.7$ but this is a
detail that does not make a difference at the qualitative level of our 
analysis here.}
Then \equ{Q} implies 
\be
\delta \simeq \sqrt{2} \frac{\sigma}{V} \, ,
\label{eq:delta_sigmav}
\ee
so $\sigma/V$ can replace $\delta$ as the variable governing the VDI.

\smallskip
The typical radius and mass of the giant clumps formed by the instability 
in a disc with respect to the corresponding disc global properties are 
\be
\frac{\Rclump}{\Rd} \simeq \frac{\pi}{4} \delta, \quad
\frac{\Mclump}{\Mcold} \simeq \frac{\pi^2}{16} \delta^2 \, .
\label{eq:Mclump}
\ee
If $\delta$ has a systematic dependence on redshift or galactic mass,
the clump mass relative to the disc mass will show correlations with these
parameters accordingly.

\subsection{Inflow within the Disk}
\label{sec:inflow}

The turbulence tends to decay on a timescale comparable to the disc
dynamical time \citep[e.g.][]{maclow99}, so it should be continuously powered
by an energy source that could stir up turbulence and maintain $\sigma$ 
at the level required for $Q \simeq 1$.
In the perturbed disc, which consists of extended transient features and 
massive compact clumps, gravitational torques drive angular momentum out
and cause mass inflow towards the centre, partly as clump migration
\citep{noguchi99,bournaud07c,dsc09} 
and partly as off-clump inflow 
\citep{gammie01,dsc09,bournaud11}.
This inflow down the potential gradient from the disc outskirts to its 
centre in turn provides the required energy for maintaining $Q \simeq 1$
\citep{krum_burkert10,burkert10,forbes12,cdg12,forbes13}. 
The gas inflow rate ${\dot M}_{\rm g,in}$
can be estimated by equating this energy gain 
and the dissipative losses of the turbulence \citep{dekel13}, 
\be
{\dot M}_{\rm g,in} V^2 \simeq \frac{\Mg \sigma^2}{\gamma \td} \, .
\label{eq:energy}
\ee
The turbulence decay timescale is the ratio of disc height to velocity
dispersion, which for $Q \sim 1$ can be expressed as $\gamma_{\rm dis} \td$,
where $\td \equiv \Rd/V$ is the global dynamical disc crossing time at the 
effective radius of the disc, and $\gamma_{\rm dis}$ is a parameter of order 
unity.  
We can think of the parameter $\gamma$ as a product of three parameters,
$\gamma  = (2/3)\gamma_{\rm g}^{-1}\gamma_{\rm \Phi}\gamma_{\rm dis}$,
where $\gamma_{\rm g}$ is the fraction of gas in the inflowing mass,
and $\gamma_\Phi V^2$ is the energy gain per unit mass when contracting from
the disc radius to the centre.
The value of $\gamma$ is thus of order unity, and it 
could be as large as a few.
We thus obtain for the inflow timescale
\be
\tinf = \frac{\Mg}{\dot{M}_{g,in}} \simeq 2\gamma\td \delta^{-2} \, .
\label{eq:tinf}
\ee
Note that $\delta$ refers to the whole cold component, including gas and 
young stars, while $\Mg$ and $\dot{M}_{g,in}$ refer to the gas only. 

\smallskip
Independent estimates based on the mechanics of driving the mass inflow by 
torques yield similar results to within a factor of 2. 
Examples of such calculations are 
(a) an estimate of the rate of energy exchange by clump encounters 
\citep[][eqs.~21 and 7]{dsc09},
(b) an estimate of the angular-momentum exchange among 
the transient perturbations in a viscous disc 
\citep[][eq.~24]{gammie01,genzel08,dsc09}, 
and (c) an estimate based on dynamical friction during clump migration.

\smallskip
One should emphasize that the inflow in the disc is a robust feature of
the instability, not limited to clump migration. This is seen in 
zoom-in AMR cosmological simulations \citep{dekel13}.
It has been argued based on theoretical estimates that the giant clumps
survive intact during their migration \citep{kd10,dk13}, but  
even if they are disrupted by stellar feedback in less than a 
migration time, as has been reproduced in simulations with unrealistically
strong radiative stellar feedback
\citep{murray10,genel12,hopkins12b,hopkins12c},
inflow at the rate comparable to \equ{tinf} is still expected
\adb{
\citep{cdg12,forbes13,dekel13,bournaud13}. 
}

\smallskip
As a note of caution, we should mention that  
the inflow rate might be modified if the turbulence is  
driven by sources other than the disc's self gravity.
For example, the cosmological streams that feed the disc may
drive non-regulated turbulence but only if the streams largely consist 
of dense clumps that allow strong coupling of the streams with the
higher-density disc \citep{dsc09,elmegreen_burkert10,gdc12}.
Stellar feedback, by supernovae and radiation pressure, may or may not be 
a major direct driver of the turbulence
\citep{dsc09,bournaud10,krum_thom13}.
The lack of correlation between velocity dispersion and star-forming regions
is evidence against it \citep{forster09,genzel11}.
However, outflows can help boosting up $\sigma$ in an indirect way,
by lowering the surface density of the remaining gas that is supposed 
to be stirred up by the given energy source \citep{gdc12}. 
In the following, we address the 
\adb{
most likely but perhaps idealized
}
case where the disc's self-gravity
is the origin of self-regulated VDI, so \equ{tinf} is valid.

\smallskip
In a cosmological steady state, where there is balance between cosmological 
accretion, star formation and VDI-driven inflow within the disc, the typical 
value is expected to be $\delta \sim 0.3$
\citep{dsc09,cdg12,forbes13,dekel13}. 
The corresponding kinematic quantity is $\sigma/V \sim 0.2$, as observed
\citep{genzel06,genzel08}.
This has been predicted analytically
\citep{dsc09,cdg12} and confirmed in cosmological simulations \citep{dekel13}.
Based on \equ{Mclump}, this implies that the clumps are giant,
with masses of a few percent of the disc mass.
The corresponding timescale for inflow in \equ{tinf} is $\sim 10 \td$,
namely 1-2 orbital times at the disc radius, or $\sim 250 \Myr$ 
at $z \sim 2$.
This is the motivation for referring to the disc instability at high redshift
as ``violent", unlike the secular, slow evolution associated with 
bar instability and spiral arms in the stellar-dominated discs at low 
redshift.

\subsection{A Critical $\delta$ for Wet Inflow}
\label{sec:wet}

The timescale for star formation in the disc roughly scales
with the free-fall time as
\be
\tsfr = \epssfr^{-1} \tff \, ,
\label{eq:tsfr1}
\ee
where the SFR efficiency is $\epssfr \sim 0.02$
\citep[][and references therein]{kdm12}.
Assuming that star formation occurs in regions that are denser than the
mean density of baryons in the disc by a factor of a few,
we adopt $\tff \simeq 0.5 \td$.

%

\smallskip
Combined with \equ{tinf}, we obtain for the wetness parameter
\be
w \simeq (4\gamma\epssfr)^{-1}\, \delta^{2} \, . 
 \label{eq:w_delta}
\ee
The condition for a wet inflow, $w>1$, thus defines a threshold cold fraction
for a wet inflow, $\delta > \delw$, with
\be
\delw \simeq 0.28\, (\gamma\, \epssfr_{0.02})^{1/2} \, .
\label{eq:delw}
\ee
It is interesting to note that this threshold is universal --- 
it does not depend systematically on galaxy mass or cosmic time.
Also interesting is the similarity between the derived value of $\delw$ 
and the typical value of $\delta \sim 0.3$ predicted for the cosmological 
steady state at high redshift,
as well as the corresponding average value of $\sigma/V \sim 0.2$ observed.

\smallskip
Note that $\delta$ can be expressed as a ratio of surface densities,
\be
\delta \simeq \frac{\Sigc}{\Sigc+\Sigh}\, ,
\ee
so $\delta$ is monotonically increasing with $\Sigc$.
We will refer to $\Sigma$ later, in comparison to observations.

\subsection{Bimodality in compactness}
\label{sec:bimodality}

Once started wet, the inflow will remain wet as long as it is within a VDI disc.
This is because as a given gas mass contracts to a smaller radius, 
$\delta$ within this radius is expected to increase, and so does $w$
according to \equ{w_delta}. To see that $\delta$ is increasing during
gas contraction, recall that it is inversely proportional to $\Mt(r)$, which 
is monotonically increasing with $r$.  
With $w$ growing, the condition for wet inflow becomes even more valid.
This implies that, as long as the VDI is valid, the wet shrinkage
is a runaway process.
The fact that $\td$ may become smaller at smaller radii scales out, as it
affects both $\tsfr$ and $\tinf$ in a similar way. 

\smallskip
This should lead to a bimodality in the surface density of the population of
star-forming galaxies. The gas in the discs in which the initial $\delta$   
is above the threshold value $\delw$ would flow in and make
$\delta$ increase away from the threshold value. 
In discs with $\delta$ below the threshold there will be 
only little or no inflow, and no change in surface density, leaving behind an
extended stellar-dominated disc. 
This should generate a gap in the galaxy population just above the threshold
value of $\delw$ (or the corresponding surface density, see below).


\subsection{Dispersion-dominated bulge}
\label{sec:dispersion}

As long as the disc is in a self-regulated VDI at $Q \simeq 1$ it has
$\sigma/V \prop \delta$. Therefore, as the gas contracts, 
and $\delta$ increases, $\sigma/V$ also increases. At some point, when
$\sigma/V$ is of order unity, this is not a disc anymore, the VDI stops, 
and the system becomes dispersion dominated.  In particular, the actual
rotation velocity falls below the circular velocity dictated by the potential
well \citep{burkert10}, and the rotation curve
does not rise at small radii to values of several hundred $\kms$.

\smallskip
For illustrative purposes, if we crudely approximate the total mass 
profile in the relevant range as a fixed isothermal sphere, 
$\Mt \propto r$, we obtain that during the gas contraction  
${\sigma}/{V} \propto \delta \propto r^{-1}$. 
Starting with a disc of $\delta \simeq 0.28$, 
where $\sigma/V \simeq \delta/\sqrt{2} \simeq 0.20$,
the contraction would proceed by a factor of 2.5 in radius till
reaching $\sigma/V \simeq 0.5$, after which the system could be considered
dispersion dominated, where standard VDI is no more valid.
We thus expect a contraction factor of a few between the original gas disc
and the compact spheroid, from a typical effective radius of a few kpc 
to one kpc, with an enhancement of the surface density by a factor of order 10.

\section{Blue nuggets and spin}
\label{sec:bn_spin}

\subsection{A Critical spin parameter}
\label{sec:spin}

Where and when do we expect VDI-driven blue nuggets, namely $\delta>\delw$?
First, at high redshift, when $\delta$ (and $\Sigc$) are high 
because the gas fraction tends to be high ($\sim 50\%$),
both observationally \citep{daddi10,tacconi10,tacconi13} 
and based on theoretical considerations \citep[e.g.][]{kd12}.
Second, in galaxies where the baryonic spin parameter happened to be 
relatively low, leading to discs with small radii and therefore higher  
$\Sigc$ and $\delta$.
Third, in galaxies where feedback has not removed a significant amount of gas,
which are likely to be the more massive galaxies. This is because
gas removal lowers $\delta$ both by reducing the gas mass and by increasing
the effective spin parameter, as the removed gas tends to come from the 
central regions where the angular momentum is low. 
One might also consider galaxies where the cosmological accretion rate 
happened to be higher than average, but the accretion-driven variations in 
$\Sigc$ \citep{forbes13} are expected to be smaller than the
variations due to the spin parameter \citep{bullock01_j}.   

\smallskip
We next use a simple toy model to express the basic parameter for VDI, $\delta$
(or $\sigma/V$), in terms of the gas fraction and the spin parameter.
We then use the distribution of spin parameter to evaluate the abundance of
blue nuggets and predict observable criteria for them, as a function of
redshift and mass.

\smallskip
We define the parameters $\lambda$ and $\md$ as the proportionality
factors between the disc baryonic mass $\Md$ and effective radius $\Re$  
and the virial mass and radius,
\be
\Md \equiv \md \Mv, \quad \Re \equiv \lambda \Rv \, .
\label{eq:md_lambda}
\ee

\smallskip
If specific angular momentum $j$ is conserved during galaxy formation, 
$\lambda$ approximates (to within a factor of 2) the halo spin parameter,
$\lambda_{\rm spin} = j/(\Rv\Vv)$ (see \se{conc}).
According to tidal-torque theory and cosmological simulations,
the halo spin parameter is roughly independent of mass and redshift.
It averages to $\la \lambda_{\rm spin} \ra \simeq 0.05$,
and it is distributed lognormally with a standard deviation of 
$\simeq 0.5$ in natural log \citep{bullock01_j}.
The effective contraction factor from the virial radius to discs 
as measured in zoom-in hydro cosmological simulations indeed averages 
to $\la \lambda \ra \simeq 0.05$ in the redshift range $z=1-4$ and for haloes
of $\Mv \sim 10^{12}\msun$ at $z=2$ \citep[][Fig.~13]{dekel13}.
We denote $\lamm \equiv \lambda/0.05$.

\smallskip
The baryon fraction in the disc is a fraction of the cosmological baryonic
fraction $\fb \simeq 0.17$, 
given that some of the gas is in the halo outside the disc 
and some has been blown away by feedback. 
As discussed below, the average value estimated from observations is 
in the ballpark of $\md \sim 0.05$.
We denote $\mdd \equiv \md/0.05$.

\smallskip
To compute $\delta$ at $\Re$ in the forming disc, 
we use $\fc$ to represent the fraction of cold baryons (gas and young
stars) in the total baryonic disc mass, 
\be
\fc \equiv \frac{M_{\rm cold}}{\Md} \, ,
\ee
so, from \equ{md_lambda}, $M_{\rm cold}(\Re) = 0.5 f_{\rm c} \md \Mv$. 
The value of $\fc$ is comparable to the gas fraction $\fg$, and may be larger
by up to a factor of 2.
We denote $\fcc \equiv f_{\rm c}/0.5$.
 
\smallskip
For illustrative purposes, we appeal again to the toy model where we crudely 
approximate the total mass profile in the relevant range of radii as an 
isothermal sphere with a flat circular velocity, namely 
$\Mt(r) \prop r$. Then, from \equ{md_lambda}, $\Mt(\Re) = \lambda \Mv$. 
We thus obtain for the relation between $\delta$ and $\lambda$ 
\be
\dellam = 0.25\, \fcc\, \mdd\, \lamm^{-1} \, .
\label{eq:dellam}
\ee

\smallskip
Note the similarity between the derived typical value 
$\dellam \sim 0.25$ for massive
galaxies at $z \sim 2$ and the estimated threshold for wet inflow, 
$\delw \sim 0.28$ of \equ{delw}.
The natural $1\sigma$ scatter of $\lambda$ is roughly from 
0.030 to 0.082 \citep{bullock01_j},
corresponding by \equ{dellam} with $\fcc \sim \mdd \sim 1$
to $\delta \sim 0.15-0.41$,
and thus by \equ{w_delta} to a wetness parameter range 
$w \sim (0.28 - 2.1) (\gamma\epssfr_{0.02})^{-1}$.
This implies that the gas-rich galaxies 
are expected to populate the whole bimodal range from wet inflow
into a compact bulge to the formation of an extended stellar disc. 

\smallskip 
Given the universal lognormal distribution of spin parameter about an average
corresponding to $\la \lambda \ra \simeq 0.05$ 
with a standard deviation of $\sigma_{{\rm ln}\lambda} \simeq 0.5$,
we can evaluate the fraction of star-forming galaxies that produce massive
compact bulges with respect to the total number of star-forming galaxies
that also includes extended discs, as a function of redshift
and mass. 
By equating $\dellam$ from \equ{dellam} with $\delw$ from \equ{delw},
we determine the threshold $\lambda$ corresponding to $w=1$, 
\be
{\lamw}_{,0.05} \simeq \Gamma\, f_{{\rm c},0.5}\, m_{{\rm d},0.05} \, ,
\label{eq:lamw}
\ee
where 
\be
\Gamma \equiv 0.89\, (\gamma\, \epssfr_{0.02})^{-1/2} \sim 1 \, .
\ee  
While this expression 
\adb{
should properly capture
}
the characteristic value of $\lamw$ and its general 
\adb{
monotonic
}
dependence on $\fc\md$, 
\adb{
the actual expression may be somewhat modified according to possible deviations 
from
}
the toy model assumed, $\Mt \prop r$.
From the distribution of $\lambda$,
the fraction of compact bulges is expected to be about one half 
for ${\lamw}=0.05$, and it should be monotonically increasing with $\lamw$.

\subsection{Fraction of Blue Nuggets}
\label{sec:fraction}

\begin{figure}
\vskip 7.1cm
\includegraphics{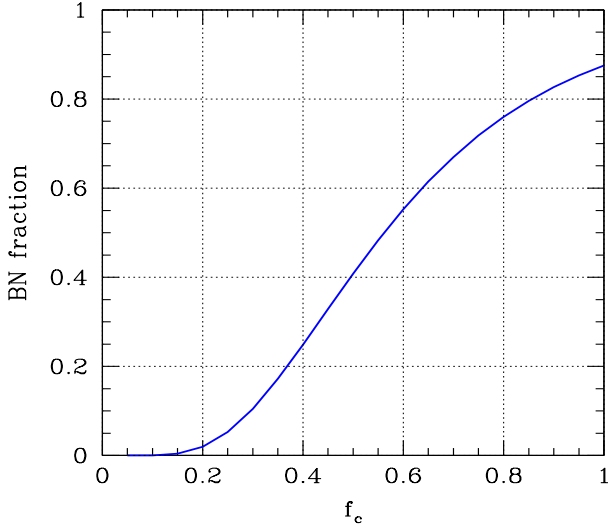}
\caption{
The fraction of star-forming galaxies that become blue nuggets
as a function of the fraction of cold mass with respect to the baryon mass,
$\fc$, (comparable to and somewhat larger 
than the gas fraction $\fg$). It is obtained from
integrating the log-normal distribution of $\lambda$ from zero to $\lamw$.
The parameter $\fc$ in the x axis actually stands for 
$\fc\, m_{d,0.05} (\gamma \epssfr_{0.02})^{-1/2}$. 
}
\label{fig:bn}
\end{figure}

\Fig{bn} shows the fraction of star-forming galaxies that become blue nuggets
as a function of $\fc\, m_{d,0.05} (\gamma \epssfr_{0.02})^{-1/2}$,
as obtained by integrating the log-normal distribution of $\lambda$ 
from zero to $\lamw$.
We read from the figure, assuming  
$\gamma \simeq \epssfr_{0.02} \simeq m_{{\rm d},0.05} \simeq 1$, 
that when $\fc \simeq 0.5$ one expects a large blue nugget fraction of 
$\sim 0.4$.
For $\fc \simeq 0.2$, the blue-nugget fraction is expected to be
drastically reduced to $\sim 0.02$, and for $\fc \simeq 0.1$ the blue-nugget 
fraction is predicted to be only $\sim 3\times 10^{-4}$.
 
\smallskip
Since the value of $\fc$ is strongly decreasing with cosmological
time (see below), we expect, at any given mass,
a significantly higher relative abundance 
of compact bulges at higher redshifts.

\smallskip
In order to evaluate the mass and redshift dependence of the blue-nugget
abundance we should come up with estimates for how the averages of the
quantities involved, namely $\lambda$, $\md$ and $\fc$, scale with mass and
redshift.
In order to obtain the qualitative behavior,
we express these crude estimates as power laws, or scaling relations.
While the analysis and results discussed so far could be considered
qualitatively robust, the following is naturally less so 
because the assumed scaling relations are rather uncertain.
We focus on the redshift range $1-3$ and the halo-mass range from
below $10^{11}\msun$ to above $10^{12.5}\msun$.

\smallskip
For massive haloes, $\Mv \geq 10^{12}\msun$, we adopt $\la \lambda \ra = 0.05$. 
For less-massive haloes we may write for the disc
\be
\la \lambda \ra = 0.05 \M12^{-\ell_m} \z3^{-\ell_z} \, ,
\label{eq:lam_av}
\ee
where $\M12\equiv \Mv/10^{12}\msun$ and $\z3\equiv (1+z)/3$.
The variation with mass and redshift may arise from feedback effects, where
the outflows tend to preferentially remove low-spin gas from the galaxy centre
and thus increase the effective $\lambda$. These effects are expected to be
stronger at lower masses and redshift, where the potential wells are shallower
\citep{ds86}.
Based on the modeling of \citet{maller02}, one can crudely
estimate $\ell_m \sim 1/3$ and $\ell_z >0$, indicating a weak dependence on
mass and redshift.

\smallskip
Based on the observational estimate of molecular gas reported in
\citet{tacconi13}, we crudely estimate
\be
\fc \simeq 0.5 \M12^{-\phi_m} \z3^{-\phi_z} \, ,
\ee
with $\phi_m \simeq 1/3$ and $\phi_z \simeq 1$.
Indeed, similar scalings with mass and redshift are predicted for the mechanism
of metallicity-dependent quenching \citep{kd12}.
Then, for the stellar fraction $\fs = 1-\fc$, near $\fc \sim 0.5$,
we can estimate
\be
\fs \simeq 0.5 \M12^{\phi_m} \z3^{\phi_z} \, .
\ee

\smallskip
In order to obtain scaling relations for $\md$, we use the 
\citet{behroozi13} estimate from observations
of $\Ms/\Mv$ as a function of $\Mv$ and $z$.
The estimated scaling relation varies depending on the relevant mass range,
as $\Ms/\Mv$ reaches a peak near $\Mv \sim 10^{12}\msun$. 
We write
\be 
{\Ms}/{\Mv} \simeq 0.025\, \M12^{\mu_m}\, \z3^{\mu_z} \, ,
\ee
and crudely approximate in two mass ranges
\be
\mu_m\!\simeq\!\mu_z\!\simeq 
\begin{cases}
 0, \   & \Mv\!\sim\!10^{11.5-12.5}\msun \\
 2/3, \ & \Mv\!\sim\!10^{11.0-12.0}\msun.
\end{cases}
\ee
Using $\md = \fs^{-1} \Ms/\Mv$ we obtain
\be
\md \simeq 0.05 \M12^{-\phi_m+\mu_m} \z3^{\phi_z-\mu_z} \, .
\ee
Then, from \equs{lamw} and \ref{eq:lam_av},
\be
{\lamw}/{\la \lambda \ra} \simeq \Gamma\, \M12^{-2\phi_m+\mu_m+\ell_m}
\z3^{2\phi_z-\mu_z+\ell_z} \, .
\ee
With the estimates for the power indices this is 
\be
\!\!\frac{\lamw}{\la \lambda \ra}\!\simeq\!
\begin{cases}
\Gamma\M12^{-1/3} \z3^2,\!     &\Mv\!\sim\!10^{11.5\!-\!12.5}\msun \\
\Gamma\M12^{1/3} \z3^{4/3},\!  &\Mv\!\sim\!10^{11.0\!-\!12.0}\msun.
\end{cases}
\label{eq:lamw_mz}
\ee
%
We see that the critical value of $\lamw$ is strongly increasing
with redshift in the whole relevant mass range. This can be translated to
a strong redshift dependence of the abundance of blue nuggets among the
star-forming galaxies. 
On the other hand, the predicted mass dependence in \equ{lamw_mz} is weak.
Any mass downsizing in quenching should have another origin,
e.g. in a downsizing of the VDI process itself or in halo-mass quenching
(see \se{conc}).

\subsection{A Critical Surface Density}
\label{sec:Sigma}

Using \equ{md_lambda} we write the stellar surface density within the disc
effective radius as
\be
\Sigs = \frac{0.5\, \fs\,\Md}{\pi \Re^2} 
= \frac{0.5\, \fs\,\md}{\pi \lambda^2} \frac{\Mv}{\Rv^2} \, .
\label{eq:Sigma1}
\ee
In the Einstein-deSitter cosmological regime, valid to a good approximation
at $z>1$, the virial quantities are related by
\be
\R100 \simeq 1.03\, \M12^{1/3} (1+z)_3^{-1} \, ,
\label{eq:vir}
\ee
where $\R100 \equiv \Rv /100\kpc$.
This gives
\begin{align}
\Sigs &\simeq (2\pi)^{-1} \times 10^9 \msun\kpc^{-2}\, \nonumber \\
&\quad \ \times \lambda_{0.05}^{-2}\, \fss\, \mdd\, \M12^{1/3}\, \z3^2 \, .
\label{eq:Sigma2}
\end{align}
Using the critical value of $\lamw$ and the scaling relations for $\fs$ and
$\md$ this becomes
\begin{align}
\Sigw &\simeq 0.2 \times 10^9 \msun \kpc^{-2} \nonumber \\
 &\quad \times \M12^{4\phi_m-2\mu_m+1/3} \z3^{-4\phi_z+2\mu_z+2} \, .
\end{align}
After the dissipative contraction, the spheroid is expected to have
contracted by a factor of a few, so the stellar surface density of the compact
spheroid is expected to be larger by a factor of $\sim 10$. 
The predicted scaling is thus
\begin{align}
\!\!\!\Sigw &\simeq 2\times 10^{9}\msun\kpc^{-2} \nonumber \\
&\quad \times
\begin{cases}
\M12^{5/3} \z3^{-2},   \!  &\Mv\!\sim\!10^{11.5-12.5}\msun\\
\M12^{1/3} \z3^{-3/2}, \!  &\Mv\!\sim\!10^{11.0-12.0}\msun.
\end{cases}
\label{eq:Sigma_crit}
\end{align}
While the magnitude of the surface density should be considered as a reliable
order-of-magnitude estimate, one should recall that the scalings with mass 
and redshift are based on uncertain scaling relations for the gas fraction,
$\md$ and $\lambda$, and should therefore be considered with a grain of salt.


\subsection{
\adb{Preliminary}
Comparison to Observations}

\adb{
Our aim in the current paper is to present the key features of the 
formation of blue nuggets through VDI, and to make qualitative predictions
based on an idealized toy model. A detailed comparison of the model to
observations is premature; it will have to wait for more quantitative
predictions, e.g., based on simulations, and for the proper observations to 
accumulate and allow detailed analysis and reliable statistical results. 
Here we briefly comment on the encouraging qualitative consistency between 
the crude theoretical predictions and the pioneering observational results
concerning blue nuggets.
}

\smallskip
The 
\adb{
predicted}
strong redshift dependence of the fraction of blue nuggets among the 
star-forming galaxies is indeed seen in the observed samples by
\citet{barro13},
which contain galaxies of stellar mass $10^{10.5-11.5}\msun$
in the redshift range $z=1-3$.
\adb{
In their Figure 8, 
comparing the blue to the grey histogram,
}
one can see that the blue-nugget fraction drops
\adb{
at a rate consistent with the predicted rate,
}
from $\sim\!40\%$ at $z\!=\!2.2\!-\!2.6$ to $\sim\!10\%$ at $z\!=\!1.0\!-\!1.4$,
and to an even smaller value by $z\!=\!0.5\!-\!1.0$.
 
\smallskip
\adb{
The star-forming galaxies at $z>2$ indeed indicate a ``bimodality" in 
compactness, consisting of a non-negligible population of compact blue nuggets 
and a large body of star-forming galaxies with low surface-densities, 
qualitatively consistent with the model predictions.
A recent analysis of CANDELS data (van der Wel et al. in preparation)
reveals a similar distribution of sizes, as a function of mass and redshift,
with the radii spread over a decade in each bin of mass and redshift, from 1 to
10 kpc.
}

\smallskip
The threshold for blue nuggets as identified by \citet{barro13} at 
$z \sim 2-3$ corresponds to an effective stellar surface density of
$\Sigs \sim 2.5\times 10^9 \msun\kpc^{-2}$.
\adb{
This is derived from their Figure 1 
(or Figure 7), 
}
using $\Ms = 10^{11}\msun$ and the
corresponding threshold value of $\Re \simeq 2.5\kpc$.
This observed threshold is consistent with the predicted threshold,
\equ{Sigma_crit}. 

\smallskip
At a redshift somewhat smaller than unity, $z=0.5-0.8$, 
and in the stellar mass range $10^{10.3-11.3}\msun$, 
\citet{cheung12} find a similar threshold for the effective surface density
of blue, green and red nuggets at $\Sigs \sim 10^9 \msun\kpc^{-2}$
\adb{
(Figure 5, bottom-left panel).
}
It is also seen as a threshold in the surface density within the
inner 1 kpc, $\Sig1 \simeq 2\times 10^9 \msun\kpc^{-2}$
\adb{
(Figure 5, bottom-left panel).
}
This is in the ball park of the prediction in \equ{Sigma_crit}.
These data do show an increase of $\Sig1$ with mass as predicted
(E.~Cheung, private communication), though the predicted redshift dependence
is not detected.
%

\smallskip
At low redshift, $z<0.075$, and in the stellar mass range 
$10^{9.75-11.25}\msun$, \citet{fang13} find
complementary evidence for blue, green and red nuggets.
By plotting the sSFR estimator as a function of $\Sig1$
in bins of stellar mass one can interpret the distribution of galaxies as
tracing evolutionary tracks. For stellar masses in the range
$10^{10.5-11.25}\msun$, the detected threshold 
\adb{
in their Figure 6
}
is at $\Sig1 \sim 3\times 10^{9} \msun\kpc^{-2}$. 
The associated value of the effective surface density can be estimated to be
a factor of a few smaller.
The overall scaling with mass 
\adb{
as reported by \citet{fang13}
}
is $\Sig1 \prop \Ms^{0.6-0.7}$,
qualitatively along the lines of the predicted scaling in \equ{Sigma_crit}.
The predicted increase of $\Sigw$ at low redshift is not seen in the data.
If the predicted scaling with redshift is valid (reflecting correct adopted
scalings of $\fc$, $\md$ and $\lambda$), the failure to reproduce it
may indicate that most of the quenching of galaxies that are now red
nuggets actually occurred at higher redshifts. 

\smallskip
Indeed,
\adb{
Figure 6 of \citet{fang13}
}
indicates downsizing, in the sense that
the relative abundance of blue and green nuggets compared to red nuggets  
is decreasing with mass. This implies that most of the more massive galaxies 
have quenched in the past while many of the less massive galaxies are quenching
right now.
The origin of this downsizing may be related to halo quenching
(\se{quenching_halo}), or to downsizing in the condition $Q \sim 1$ for VDI. 


\smallskip
Independently, based on measurements of cold gas at low redshift,
\citet{kauffmann12} report a threshold effective surface density for
gas-poor galaxies at $\Sigs \simeq 10^9 \msun \kpc^{-2}$ 
for $\Ms \sim 10^{11}\msun$, with a scaling $\Sigs \prop \Ms^{0.4}$
\adb{
(Figures 9 and 10, top-left panels).
}
These are consistent with the findings of \citet{fang13}, and with the
theoretical prediction, \equ{Sigma_crit}.

\smallskip
\adb{
\citet{barro13b} have recently studied the structural properties and stellar 
populations of 45 massive blue nuggets from the CANDELS survey
at $z=2-3$, and provided further 
evidence that these galaxies are consistent with being the
descendants of the more extended star-forming discs
and the progenitors of the red nuggets at $z \sim 2$.
In particular, the indicated contraction factors from discs to blue nuggets, 
and the more spheroidal morphologies of the latter, are consistent with the
model predictions for a contraction factor of a few and for the system becoming
dispersion dominated during the contraction.
}

\section{Quenching: Blue to Red Nuggets}
\label{sec:quenching}

\adb{
The creation of compact, dispersion-dominated blue nuggets allows the
development into compact red nuggets of similar structural properties.
Furthermore,
the high SFR induced by the high gas density provides natural conditions
for the subsequent quenching of star formation necessary for the development
into passive red nuggets. 
We provide here a discussion of the possible quenching mechanism,
trying to distinguish between internal mechanisms related to the central
structure of the galaxies and external mechanisms associated with the
circum-galactic medium (CGM) in the dark-matter halo.
We then discuss the possible association of these quenching mechanisms
with fast and slow evolution modes, both in the quenching and the
compactification phases, and connect these modes with the disc wetness 
parameter and the galaxy spin parameter as introduced in our analysis above. 
}

\subsection{Internal Quenching}
\label{sec:quenching_bulge}

A compact gas bulge with a high $\Sigg$ naturally gives rise to a central phase
of high SFR. This could cause a drastic decrease in $\Sigg$ by gas consumption
into stars as well as gas outflow due to stellar feedback, and thus lead to
rapid quenching, 
\adb{
possibly 
}
over $\sim\!1\Gyr$ or less. 
Similarly, the dissipative inflow leads to a rapid growth of
the central black hole \citep{bournaud11}, 
and the quenching could be a result of gas removal by AGN feedback
\citep[though perhaps not immediately][]{gabor13}.

\smallskip
The development of
a compact massive bulge could also lead to quenching by stabilizing the VDI,
where the Toomre $Q$ parameter is driven to values above unity.
Recalling that $Q \prop \Omega \sigma / \Sigc$, \equ{Q}, 
\adb{
we notice that Q-quenching can happen in three different ways, 
or in a combined way,
}
as follows:
\rf\bul
Morphological quenching \citep{martig09}, where the growth of total central 
density is expressed as an increase in $\Omega$. 
There is observational indication for such an effect
\citep{saintonge12,crocker12,fang13,martig13},
\adb{
and recent evidence in spectroscopically observed $z \sim 2$ galaxies 
\citep{genzel13}.
}
\rf\bul
A drop in $\Sigg$, and thus in $\Sigc$, by star formation and outflows. 
In this case the mass gradually becomes dominated by stars, 
whose velocity dispersion becomes higher than that of the gas,  
which increases the effective $Q$ further. The consequent shutdown of the
VDI-driven inflow makes $\Sigg$ even lower, such that $Q$ rises in a runaway
process. This is demonstrated in analytic calculations and one-dimensional
simulations \citep{cdg12,forbes13}, where $\Omega$ is kept fixed. 
\rf\bul
An increase in the gas $\sigma$ as a result of stellar or AGN feedback
\citep{krum_thom13}, which quenches further star formation.

\subsection{Halo Quenching}
\label{sec:quenching_halo}

The observed distribution of galaxies 
in the plane of specific SFR (sSFR) versus $\Sigs$
has an ``L" shape, defined by a star-forming side of varying $\Sigs$ 
at a high sFSR and a compact side of varying sSFR at a high $\Sigs$,
along which one envisions shrinkage and quenching respectively.
In particular, the compact side reveals that
a compact galaxy could be either star forming or quenched or in
transition between the two \citep{kauffmann12,cheung12,fang13,barro13}. 
This indicates that the compactness is only a necessary
condition for quenching, not sufficient, so a complementary second mechanism is 
required for quenching. 
Independently, the shutdown of cold accretion into the halo centre
by a hot CGM in the halo is clearly necessary 
for the long-term maintenance of quenching. 
Such a halo quenching could be the required second mechanism.
A study of SDSS galaxies at low redshift indeed indicates that  
both compactness and a massive halo are required for a high quenched fraction
of galaxies (J. Woo et al. in preparation).
\adb{
This work also indicates that 
}
halo quenching is likely to be a slow mode of quenching, lasting a few Gyr,
as the cold gas reservoir in the disc is slowly exhausted and the SFR is
fading out.

\smallskip
The CGM is likely to be heated by a stable virial shock once the halo
grows above a critical mass on the order of $\sim 10^{12}\msun$ 
\citep{bd03,keres05,db06,keres09}.
While at $z>2$ narrow cold streams penetrate the hot medium and keep the SFR 
high, at lower redshifts the penetration is halted once the medium 
has been shock heated. 
Besides halting the cold streams, the presence of a hot medium improves
the coupling between AGN-driven wind and the accreting gas, and thus makes the
AGN feedback more effective in quenching the SFR \citep{db06}.

\smallskip
In even more massive haloes, $\Mv \geq 10^{13}\msun$, the gravitational power
associated with the gas streaming into the halo can provide the heating that
 compensate for the 
cooling rate. If the accretion is appropriately clumpy, or if the streams
couple to the medium, this gravitational heating can cause proper 
quenching at the halo centre 
\citep{db08,khochfar08}.
In addition, the dynamical friction that acts on subhaloes or satellite 
galaxies as they spiral into the central galaxy extracts energy (and angular 
momentum) from their orbits and can thus heat up the medium
\citep{elzant+kam04}.

\subsection{Fast and Slow modes}

The relative role of internal quenching and halo quenching 
may determine whether the quenching is fast or slow.
As eluded to above,
this can be decided by the degree of wetness in the original disc
\adb{
(which is a function of the galaxy spin parameter).
}

\smallskip
In the fast mode,
a disc that forms with $w \gg 1$ would form stars at a 
high rate and move quickly along the blue sequence towards larger stellar 
masses and higher central surface densities.
Once the gas central density is high, a rapid starburst would lead to
rapid quenching. The galaxies in this fast mode are to be found in the blue
side of the blue sequence, and in the high-$\Sigs$ side of the green valley,
along the external side of the L-shape track.
By the time the galaxy is on the red sequence, the quenching is maintained
if the halo is already above the critical mass so the CGM is hot. 
This fast mode is expected to be dominant once the gas fraction is high, namely
preferentially at high redshift. 

\smallskip
In the slow mode,
a disc that has $w \gsim 1$ would form stars in a moderate rate and spend more
time on the red sequence, as it slowly evolves toward higher stellar masses
and higher central densities. In parallel to slowly developing a compact core,
the haloes are growing and becoming massive enough to support a virial shock 
that causes slow quenching towards the red sequence.
The slow-mode galaxies are to be found in the red side of the blue sequence,
and then on the low-$\Sigs$ side of the green valley, along an interior
L-shape track. The halo masses are expected to be somewhat more massive on the 
slow-mode track.
 
\smallskip
Galaxies with $w < 1$ would form stars in a disc over on a long
timescale, leaving behind an extended disc. They may remain star forming,
like the Milky Way, or eventually quench by the halo when it is sufficiency
massive.

\subsection{Downsizing in Quenching}
\label{sec:downsizing}

While the above analysis of the consequences of wetness does not seem
to explicitly predict that more massive galaxies are quenched earlier,
there are several other mechanisms that naturally lead to downsizing of 
quenching \citep[also mentioned in][]{bournaud11}, 
\adb{
which we should briefly discuss here for completeness.
}.

\smallskip
First, such a downsizing is a natural prediction of halo quenching
\citep{db06,cattaneo08},
where more massive haloes tend to cross the threshold mass for virial shock 
heating earlier.
In turn, a lower-mass threshold for the onset of star formation helps
the massive galaxies form stars early, which by itself leads to downsizing
\citep{neistein06,bouche10,kd12}.

\smallskip
Second, there is a natural downsizing in the conditions for VDI, which could 
translate to downsizing in the formation of blue nuggets, and then red 
nuggets. The Toomre $Q$ parameter tends to be lower for less massive
galaxies, e.g., because of the higher cold gas fraction that is maintained
in low-mass galaxies to lower redshifts \citep[e.g.][]{kannappan04,tacconi13}. 
This could be a natural result of
stronger non-ejective feedback in low-mass galaxies \citep{kd12}.
Another reason for quenching in more massive galaxies is the
more efficient growth of bulge-to-disc ratio in more massive galaxies,
leading to earlier morphological quenching \citep{martig09}.

\subsection{\adb{Preliminary Comparison to Observations}}

\smallskip
Evidence for the dominance of a fast quenching mode at high redshifts
is provided by the rather rapid appearance of the red sequence
during a period shorter than $\sim\!0.7\Gyr$ between
$z \sim 2.8$ and $z \sim 2.2$ \citep[][Figure 2]{barro13}.
An observational indication for the presence of a minor fast mode of
quenching at low redshifts is found in green-valley SDSS galaxies in transition
from blue to red  (H. Yesuf et al. in preparation).
A small fraction of the green-valley galaxies are identified as post-starburst
galaxies, which implies that they are quenching rather rapidly,
while the rest are quenching more slowly.
Indeed, these fast-mode galaxies tend to be of higher surface density and lower
halo mass.

\smallskip
\adb{
An intrinsic difficulty in detecting the fast mode of evolution arises 
from the fact that the blue-nugget phase with high SFR may be rather short, 
implying a low observable abundance of blue nuggets. The abundance is expected
to be low even if a significant fraction of the star-forming galaxies have
shrunk to blue nuggets and most of the red nuggets have passed through
this phase, as expected at high redshift.
Detecting such galaxies on the blue side of the green valley as
post-starburst galaxies may be more feasible.
}

\smallskip
\adb{
Studies of the global shapes of $z \sim 2$ red nuggets indicate that they 
tend to be flattened spheroids, sometimes characterized as ``disk-like" or
``disk-dominated" \citep{vanderwel11,chang13}.
This is consistent with the idea that the progenitor blue nuggets
formed by VDI-driven inflow from rotating discs,
and that the subsequent quenching was not associated with a dramatic event that
is likely to change the global shape, like a major merger.
}

\section{Conclusion and Discussion}
\label{sec:conc}

\begin{figure}
\vskip 6.3cm
\includegraphics{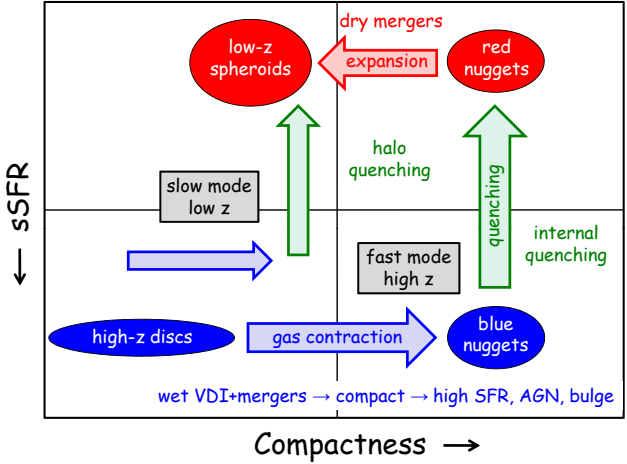}
\caption{
\adb{
A schematic diagram illustrating the scenario addressed in this paper.
The fast mode, likely dominant at high redshift, involves 
VDI-driven wet gas contraction of the denser high-redshift discs into compact, 
gaseous, star-forming blue nuggets, 
followed by quenching into compact, passive, stellar red nuggets. 
The blue-nugget compactness triggers the quenching by internal processes 
including stellar and AGN feedback and Q-quenching, to be maintained by 
hot-halo quenching. 
The red nuggets eventually de-compactify by dry mergers into the low-z 
spheroids.
A slower mode, potentially dominant at low redshift, may involve 
medium-wetness discs that contract slowly and possibly quench by halo only, 
without passing through a blue-nugget phase.
High angular momentum galaxies form extended stellar-dominated discs (not
illustrated here).
}
}
\label{fig:fig_nuggets}
\end{figure}

Given that the high-redshift compact ellipticals  
\adb{are likely to have formed}
by dissipative gas inflow toward the galaxy centre, we investigated the
\adb{likely}
main channel for this process at high redshift --- the gas contraction 
naturally associated with VDI.
\adb{ \Fig{fig_nuggets} schematically illusrates the main components of the
proposed scenario.}

\smallskip
Our main conclusions are as follows:
\rf\bul 
Given a disc in a VDI phase, a necessary condition for wet inflow is that
the timescale for star formation is longer than the timescale for inflow,
namely that the wetness parameter, $w=\tsfr/\tinf$, is larger than unity.
\rf\bul
This condition is fulfilled once the cold-to-total mass ratio within the disc
radius is above a threshold value $\delw \simeq 0.28$.
This happens to be comparable to the value obtained by discs in a cosmological
steady state.
\rf\bul
For a cold-to-baryonic mass ratio $\fc \sim 0.5$, 
and a baryonic fraction within the virial radius of $\md \sim 0.05$, 
both being approximately valid at $z \sim 2$, 
the threshold at $w=1$ corresponds to a spin parameter of $\lamw \simeq 0.05$,
which is near the average spin parameter for haloes. Given the lognormal
distribution of spin parameters \citep{bullock01_j}, this implies that
about one half of the star-forming galaxies are expected to contract to blue 
nuggets, while the rest form extended stellar discs.
\rf\bul
Blue nuggets are expected to be much less frequent at low redshifts,
\adb{both relative to the diffuse star-forming galaxies, 
and the quenched galaxies.}
\rf\bul
The star-forming galaxies at high $z$ are expected to show a corresponding
bimodality in effective stellar surface density about a value 
$\Sigs \sim 10^{9}\msun\kpc^{-2}$.
This critical value is expected to be weakly increasing with mass.
\rf\bul
The blue nuggets quench to red nuggets by mechanisms of two types.
Internal quenching can be associated with starbursts, stellar or AGN feedback,
and Q-quenching. It can serve for triggering the quenching,
and may be responsible for a fast transition from blue to red nuggets,
especially valid at high redshift.
Halo quenching can shut down the accretion onto the galaxy due to 
virial shock heating or gravitational-infall heating
once the halo grows above a critical mass of $\sim 10^{12}\msun$.
It can serve for maintenance of the quenched phase, and may be responsible
for a slow mode of quenching, valid mostly at low redshifts.
\rf\bul
These predictions are largely confirmed in hydro-cosmological simulations
(Zolotov et al. in preparation)
and are demonstrated using semi-analytic simulations (Porter et al. in
preparation).
\adb{
Both show evolutionary tracks similar to those predicted by the toy model
and indicated from the observations, where gaseous disks shrink to 
blue nuggets, which quench to red nuggets, and then gradually grow in size
by dry mergers.
}
\rf\bul
The predictions are qualitatively consistent with the results obtained from
observations of the star-forming and quenched galaxy populations, 
focusing on the blue and red nuggets at $z<1$
\citep{cheung12,kauffmann12,fang13}
and at $z=1-3$
\adb{
\citep[][Bruce et al. in preparation]{barro13,kaviraj13a,kaviraj13b,
lee13,williams13,barro13b}.
}

\smallskip
In the context of the role of the wetness parameter $w=\tsfr/\tinf$
in determining the galaxy type,
we recall that early theories of galaxy formation have proposed that 
the formation of discs and spheroids can be determined by the timescale for
star formation with respect to the dynamical time for the galaxy buildup
\citep{larson76}. 
However, there the buildup timescale referred to the global infall into the 
galaxy, while here it is specifically the VDI-driven inflow within the disc.
The two timescales are indirectly related, as the cosmological instreaming is
the driver of VDI, which in turn generates the inflow within the disc,
but they are not necessarily the same.
In particular, the cosmological accretion time at $z>1$ is proportional to 
$(1+z)^{-5/2}$ \citep[e.g.][]{neistein08b,dekel13},
while the disc inflow rate is proportional to the dynamical time $\td$,
which scales like the Hubble time, $t \propto (1+z)^{-3/2}$.

\smallskip
Similarly, the idea that the spin parameter can distinguish between spheroids
and discs has been around for a long time 
\citep{fe80,blum84,mmw98,dutton07}.
However, this idea had difficulties in the context of the common wisdom
that spheroids form by major mergers of discs, which commonly generate 
high-spin systems.
The role of the spin parameter in determining the fate of a galaxy becomes
clearer in the context of spheroid formation by VDI.

\smallskip
The estimates of the blue-nugget abundance were based on the assumption
that the distribution of specific angular momentum of the baryons is similar to
that of the total halo matter. 
This is only a crude approximation. On one hand, simulations indicate that the
specific angular momentum in the incoming baryon streams is higher 
than that of the dark-matter halo. 
On the other hand, the baryons exchange angular momentum with the dark matter
due to strong torques as they stream in 
\citep[][M. Danovich et al. in preparation]{pichon11,kimm11,tillson12,codis12,
danovich12,stewart13}.
The quantitative estimates should therefore be trusted to an accuracy of
no more than a factor of 2.

\smallskip
The scenario addressed here predicts the formation of blue nuggets as
dispersion-dominated star-forming systems, with $\sigma/V\!\geq\!0.5$.
This reminds us of the observed population of dispersion-dominated systems
\citep{law12,newman13}.
However, while the latter tend to be less massive and of lower metallicity,
the blue nuggets discussed here could be massive and of relatively
high metallicity.
It thus seems plausible that there are two types of dispersion-dominated
star-forming systems,
one that represents the early stages of evolution of a VDI disc,
and the other that arises as a product of VDI --- the blue nuggets.

\smallskip
The scenario discussed above also predicts the high-redshift presence of young 
blue nuggets surrounded by star-forming clumpy discs or rings. 
There are already possible detections
of such cases \citep[][Figure 9]{elmegreen_d05,bournaud11b}.

\smallskip
The low-redshift descendants of the red nuggets resulting from blue nuggets 
that originated from VDI discs are expected to be rotating spheroids.
The predicted systems resemble most of the present day elliptical galaxies, 
which show rapid rotation within their effective radii \citep{emsellem11}, 
dropping to slow rotation velocities at a few effective radii 
(Romanowsky et al. in preparation).
The giant non-rotating ellipticals must have formed by subsequent dry mergers.

\smallskip
\adb{
Our toy model has not addressed the detailed star formation and stellar 
evolution during the compactification into blue nuggets and quenching to red 
nuggets, so we do not attempt here to make quantitative predictions for the
abundances of galaxies in the different stages of evolution, 
or for the detailed properties of the fast and slow modes of evolution.
This will be done utilizing cosmological simulations, where mergers also
contribute to the evolution, and the evolution of the stellar population,
including SFR and feedback, is explicitly followed 
(e.g. Zolotov et al. in preparation).
The simulations will allow comparison of theory to observations of the 
evolutionary modes at low and especially high redshift, as relevant 
observations accumulate.
}

\section*{Acknowledgments}

We acknowledge stimulating discussions with Frederic Bournaud, Sandy Faber,
David Koo, Reinhard Genzel, Mark Krumholz, Joel Primack, and Joanna Woo. 
This work was supported 
by GIF grant G-1052-104.7/2009 and by a DIP grant.
The work of AD was supported by ISF grant 24/12,
by NSF grant AST-1010033,
and by the I-CORE Program of the PBC and The ISF grant 1829/12.
AB acknowledges financial support
by the Cluster of Excellence ``Origin and Structure of the Universe".



\bibliographystyle{mn2e}
\bibliography{dekel_nuggets}

\begin{thebibliography}{107}
\expandafter\ifx\csname natexlab\endcsname\relax\def\natexlab#1{#1}\fi

\bibitem[{{Agertz}, {Teyssier} \& {Moore}(2009){Agertz}, {Teyssier}, \&
  {Moore}}]{agertz09}
{Agertz} O., {Teyssier} R., {Moore} B., 2009, \mnras, 397, L64

\bibitem[{{Barnes} \& {Hernquist}(1991)}]{barnes91}
{Barnes} J.~E., {Hernquist} L.~E., 1991, \apjl, 370, L65

\bibitem[{{Barro} {et~al}\mbox{.}(2013{\natexlab{a}}){Barro}, {Faber},
  {Perez-Gonzalez}, {Koo}, {Williams}, {Kocevski}, {...}, {Dekel}, \& {et
  al.,}}]{barro13}
{Barro} G. {et~al.}, 2013{\natexlab{a}}, \apj, 765, 104

\bibitem[{{Barro} {et~al}\mbox{.}(2013{\natexlab{b}}){Barro}, {Faber},
  {Perez-Gonzalez}, {Trump}, {Koo}, {Wuyts}, {Guo}, {Dekel}, \& {et
  al.,}}]{barro13b}
{Barro} G. {et~al.}, 2013{\natexlab{b}}, arXiv:1311:0000

\bibitem[{{Behroozi}, {Wechsler} \& {Conroy}(2013){Behroozi}, {Wechsler}, \&
  {Conroy}}]{behroozi13}
{Behroozi} P.~S., {Wechsler} R.~H., {Conroy} C., 2013, \apjl, 762, L31

\bibitem[{{Birnboim} \& {Dekel}(2003)}]{bd03}
{Birnboim} Y., {Dekel} A., 2003, \mnras, 345, 349

\bibitem[{{Blumenthal} {et~al}\mbox{.}(1984){Blumenthal}, {Faber}, {Primack},
  \& {Rees}}]{blum84}
{Blumenthal} G.~R., {Faber} S.~M., {Primack} J.~R., {Rees} M.~J., 1984, \nat,
  311, 517

\bibitem[{{Bouch{\'e}} {et~al}\mbox{.}(2010){Bouch{\'e}}, {Dekel}, {Genzel},
  {Genel}, {Cresci}, {F{\"o}rster Schreiber}, {Shapiro}, {Davies}, \&
  {Tacconi}}]{bouche10}
{Bouch{\'e}} N. {et~al.}, 2010, \apj, 718, 1001

\bibitem[{{Bournaud} {et~al}\mbox{.}(2011{\natexlab{a}}){Bournaud}, {Chapon},
  {Teyssier}, {Powell}, {Elmegreen}, {Elmegreen}, {Duc}, {Contini}, {Epinat},
  \& {Shapiro}}]{bournaud11b}
{Bournaud} F. {et~al.}, 2011{\natexlab{a}}, \apj, 730, 4

\bibitem[{{Bournaud} {et~al}\mbox{.}(2008){Bournaud}, {Daddi}, {Elmegreen},
  {Elmegreen}, {Nesvadba}, {Vanzella}, {di Matteo}, {Le Tiran}, \& {et
  al.,}}]{bournaud08}
{Bournaud} F. {et~al.}, 2008, \aap, 486, 741

\bibitem[{{Bournaud} {et~al}\mbox{.}(2011{\natexlab{b}}){Bournaud}, {Dekel},
  {Teyssier}, {Cacciato}, {Daddi}, {Juneau}, \& {Shankar}}]{bournaud11}
{Bournaud} F., {Dekel} A., {Teyssier} R., {Cacciato} M., {Daddi} E., {Juneau}
  S., {Shankar} F., 2011{\natexlab{b}}, \apjl, 741, L33

\bibitem[{{Bournaud} \& {Elmegreen}(2009)}]{bournaud09}
{Bournaud} F., {Elmegreen} B.~G., 2009, \apjl, 694, L158

\bibitem[{{Bournaud}, {Elmegreen} \& {Elmegreen}(2007){Bournaud}, {Elmegreen},
  \& {Elmegreen}}]{bournaud07c}
{Bournaud} F., {Elmegreen} B.~G., {Elmegreen} D.~M., 2007, \apj, 670, 237

\bibitem[{{Bournaud} {et~al}\mbox{.}(2010){Bournaud}, {Elmegreen}, {Teyssier},
  {Block}, \& {Puerari}}]{bournaud10}
{Bournaud} F., {Elmegreen} B.~G., {Teyssier} R., {Block} D.~L., {Puerari} I.,
  2010, \mnras, 409, 1088

\bibitem[{{Bournaud} {et~al}\mbox{.}(2013){Bournaud}, {Perret}, {Renaud},
  {Dekel}, {Elmegreen}, {Elmegreen}, {Teyssier}, \& {et al.,}}]{bournaud13}
{Bournaud} F., {Perret} V., {Renaud} F., {Dekel} A., {Elmegreen} B.~G.,
  {Elmegreen} D.~M., {Teyssier} R., {et al.,}, 2013, arXiv:1307.7136

\bibitem[{{Bullock} {et~al}\mbox{.}(2001){Bullock}, {Dekel}, {Kolatt},
  {Kravtsov}, {Klypin}, {Porciani}, \& {Primack}}]{bullock01_j}
{Bullock} J.~S., {Dekel} A., {Kolatt} T.~S., {Kravtsov} A.~V., {Klypin} A.~A.,
  {Porciani} C., {Primack} J.~R., 2001, \apj, 555, 240

\bibitem[{{Burkert} {et~al}\mbox{.}(2010){Burkert}, {Genzel}, {Bouch{\'e}},
  {Cresci}, {Khochfar}, {Sommer-Larsen}, {Sternberg}, \& {et al.,}}]{burkert10}
{Burkert} A., {Genzel} R., {Bouch{\'e}} N., {Cresci} G., {Khochfar} S.,
  {Sommer-Larsen} J., {Sternberg} A., {et al.,}, 2010, \apj, 725, 2324

\bibitem[{{Cacciato}, {Dekel} \& {Genel}(2012){Cacciato}, {Dekel}, \&
  {Genel}}]{cdg12}
{Cacciato} M., {Dekel} A., {Genel} S., 2012, \mnras, 421, 818

\bibitem[{{Cattaneo} {et~al}\mbox{.}(2008){Cattaneo}, {Dekel}, {Faber}, \&
  {Guiderdoni}}]{cattaneo08}
{Cattaneo} A., {Dekel} A., {Faber} S.~M., {Guiderdoni} B., 2008, \mnras, 389,
  567

\bibitem[{{Cattaneo} {et~al}\mbox{.}(2009){Cattaneo}, {Faber}, {Binney},
  {Dekel}, {Kormendy}, \& {Mushotzky}}]{cattaneo09}
{Cattaneo} A., {Faber} S.~M., {Binney} J., {Dekel} A., {Kormendy} J.,
  {Mushotzky} R., 2009, \nat, 460, 213

\bibitem[{{Cattaneo} {et~al}\mbox{.}(2013){Cattaneo}, {Woo}, {Dekel}, \&
  {Faber}}]{cattaneo13}
{Cattaneo} A., {Woo} J., {Dekel} A., {Faber} S.~M., 2013, \mnras, 430, 686

\bibitem[{{Ceverino}, {Dekel} \& {Bournaud}(2010){Ceverino}, {Dekel}, \&
  {Bournaud}}]{cdb10}
{Ceverino} D., {Dekel} A., {Bournaud} F., 2010, \mnras, 404, 2151

\bibitem[{{Ceverino} {et~al}\mbox{.}(2012){Ceverino}, {Dekel}, {Mandelker},
  {Bournaud}, {Burkert}, {Genzel}, \& {Primack}}]{ceverino12}
{Ceverino} D., {Dekel} A., {Mandelker} N., {Bournaud} F., {Burkert} A.,
  {Genzel} R., {Primack} J., 2012, \mnras,

\bibitem[{{Chang} {et~al}\mbox{.}(2013){Chang}, {van der Wel}, {Rix}, {Holden},
  {Bell}, {McGrath}, {Wuyts}, \& {et al.}}]{chang13}
{Chang} Y.-Y., {van der Wel} A., {Rix} H.-W., {Holden} B., {Bell} E.~F.,
  {McGrath} E.~J., {Wuyts} S., {et al.}, 2013, \apj, 773, 149

\bibitem[{{Cheung} {et~al}\mbox{.}(2012){Cheung}, {Faber}, {Koo}, {Dutton},
  {Simard}, {McGrath}, {Huang}, {Bell}, {Dekel}, \& {et al.,}}]{cheung12}
{Cheung} E. {et~al.}, 2012, \apj, 760, 131

\bibitem[{{Ciotti} \& {Ostriker}(2007)}]{ciotti07}
{Ciotti} L., {Ostriker} J.~P., 2007, astro-ph/0703057

\bibitem[{{Codis} {et~al}\mbox{.}(2012){Codis}, {Pichon}, {Devriendt}, {Slyz},
  {Pogosyan}, {Dubois}, \& {Sousbie}}]{codis12}
{Codis} S., {Pichon} C., {Devriendt} J., {Slyz} A., {Pogosyan} D., {Dubois} Y.,
  {Sousbie} T., 2012, \mnras, 427, 3320

\bibitem[{{Crocker} {et~al}\mbox{.}(2012){Crocker}, {Krips}, {Bureau}, {Young},
  {Davis}, {Bayet}, {Alatalo}, \& {et al.}}]{crocker12}
{Crocker} A., {Krips} M., {Bureau} M., {Young} L.~M., {Davis} T.~A., {Bayet}
  E., {Alatalo} K., {et al.}, 2012, \mnras, 421, 1298

\bibitem[{{Daddi} {et~al}\mbox{.}(2010){Daddi}, {Bournaud}, {Walter},
  {Dannerbauer}, {Carilli}, {Dickinson}, {Elbaz}, {Morrison}, \& {et
  al.}}]{daddi10}
{Daddi} E. {et~al.}, 2010, \apj, 713, 686

\bibitem[{{Damjanov} {et~al}\mbox{.}(2011){Damjanov}, {Abraham}, {Glazebrook},
  {McCarthy}, {Caris}, {Carlberg}, {Chen}, \& {et al.}}]{damjanov11}
{Damjanov} I., {Abraham} R.~G., {Glazebrook} K., {McCarthy} P.~J., {Caris} E.,
  {Carlberg} R.~G., {Chen} H.-W., {et al.}, 2011, \apjl, 739, L44

\bibitem[{{Damjanov} {et~al}\mbox{.}(2009){Damjanov}, {McCarthy}, {Abraham},
  {Glazebrook}, {Yan}, {Mentuch}, {Le Borgne}, \& {et al.}}]{damjanov09}
{Damjanov} I., {McCarthy} P.~J., {Abraham} R.~G., {Glazebrook} K., {Yan} H.,
  {Mentuch} E., {Le Borgne} D., {et al.}, 2009, \apj, 695, 101

\bibitem[{{Danovich} {et~al}\mbox{.}(2012){Danovich}, {Dekel}, {Hahn}, \&
  {Teyssier}}]{danovich12}
{Danovich} M., {Dekel} A., {Hahn} O., {Teyssier} R., 2012, \mnras, 422, 1732

\bibitem[{{Dekel} \& {Birnboim}(2006)}]{db06}
{Dekel} A., {Birnboim} Y., 2006, \mnras, 368, 2

\bibitem[{{Dekel} \& {Birnboim}(2008)}]{db08}
{Dekel} A., {Birnboim} Y., 2008, \mnras, 383, 119

\bibitem[{{Dekel} {et~al}\mbox{.}(2009){Dekel}, {Birnboim}, {Engel},
  {Freundlich}, {Goerdt}, {Mumcuoglu}, {Neistein}, {Pichon}, {Teyssier}, \&
  {Zinger}}]{dekel09}
{Dekel} A. {et~al.}, 2009, \nat, 457, 451

\bibitem[{{Dekel} \& {Krumholz}(2013)}]{dk13}
{Dekel} A., {Krumholz} M.~R., 2013, \mnras, 432, 455

\bibitem[{{Dekel}, {Sari} \& {Ceverino}(2009){Dekel}, {Sari}, \&
  {Ceverino}}]{dsc09}
{Dekel} A., {Sari} R., {Ceverino} D., 2009, \apj, 703, 785

\bibitem[{{Dekel} \& {Silk}(1986)}]{ds86}
{Dekel} A., {Silk} J., 1986, \apj, 303, 39

\bibitem[{{Dekel} {et~al}\mbox{.}(2013){Dekel}, {Zolotov}, {Tweed}, {Cacciato},
  {Ceverino}, \& {Primack}}]{dekel13}
{Dekel} A., {Zolotov} A., {Tweed} D., {Cacciato} M., {Ceverino} D., {Primack}
  J.~R., 2013, \mnras, 435, 999

\bibitem[{{Dutton} {et~al}\mbox{.}(2007){Dutton}, {van den Bosch}, {Dekel}, \&
  {Courteau}}]{dutton07}
{Dutton} A.~A., {van den Bosch} F.~C., {Dekel} A., {Courteau} S., 2007, \apj,
  654, 27

\bibitem[{{El-Zant}, {Kim} \& {Kamionkowski}(2004){El-Zant}, {Kim}, \&
  {Kamionkowski}}]{elzant+kam04}
{El-Zant} A.~A., {Kim} W.-T., {Kamionkowski} M., 2004, \mnras, 354, 169

\bibitem[{{Elmegreen} \& {Burkert}(2010)}]{elmegreen_burkert10}
{Elmegreen} B.~G., {Burkert} A., 2010, \apj, 712, 294

\bibitem[{{Elmegreen}, {Zhang} \& {Hunter}(2012){Elmegreen}, {Zhang}, \&
  {Hunter}}]{elmegreen12}
{Elmegreen} B.~G., {Zhang} H.-X., {Hunter} D.~A., 2012, \apj, 747, 105

\bibitem[{{Elmegreen}, {Elmegreen} \& {Ferguson}(2005){Elmegreen}, {Elmegreen},
  \& {Ferguson}}]{elmegreen_d05}
{Elmegreen} D.~M., {Elmegreen} B.~G., {Ferguson} T.~E., 2005, \apjl, 623, L71

\bibitem[{{Emsellem} {et~al}\mbox{.}(2011){Emsellem}, {Cappellari},
  {Krajnovi{\'c}}, {Alatalo}, {Blitz}, {Bois}, {Bournaud}, \& {et
  al.}}]{emsellem11}
{Emsellem} E., {Cappellari} M., {Krajnovi{\'c}} D., {Alatalo} K., {Blitz} L.,
  {Bois} M., {Bournaud} F., {et al.}, 2011, \mnras, 414, 888

\bibitem[{{Fabian}(2012)}]{fabian12}
{Fabian} A.~C., 2012, \araa, 50, 455

\bibitem[{{Fall} \& {Efstathiou}(1980)}]{fe80}
{Fall} S.~M., {Efstathiou} G., 1980, \mnras, 193, 189

\bibitem[{{Fang} {et~al}\mbox{.}(2013){Fang}, {Faber}, {Koo}, \&
  {Dekel}}]{fang13}
{Fang} J.~J., {Faber} S.~M., {Koo} D.~C., {Dekel} A., 2013, \apj, 776, 63

\bibitem[{{Forbes}, {Krumholz} \& {Burkert}(2012){Forbes}, {Krumholz}, \&
  {Burkert}}]{forbes12}
{Forbes} J., {Krumholz} M., {Burkert} A., 2012, \apj, 754, 48

\bibitem[{{Forbes} {et~al}\mbox{.}(2013){Forbes}, {Krumholz}, {Burkert}, \&
  {Dekel}}]{forbes13}
{Forbes} J.~C., {Krumholz} M.~R., {Burkert} A., {Dekel} A., 2013,
  arXiv:1305.2925

\bibitem[{{Forster Schreiber} {et~al}\mbox{.}(2009){Forster Schreiber},
  {Genzel}, {Bouche}, {Cresci}, {Davies}, {Buschkamp}, {Shapiro}, {Tacconi}, \&
  {et al.,}}]{forster09}
{Forster Schreiber} N.~M. {et~al.}, 2009, \apj, 706, 1364

\bibitem[{{Gabor} \& {Bournaud}(2013)}]{gabor13}
{Gabor} J.~M., {Bournaud} F., 2013, \mnras, 434, 606

\bibitem[{{Gammie}(2001)}]{gammie01}
{Gammie} C.~F., 2001, \apj, 553, 174

\bibitem[{{Genel}, {Dekel} \& {Cacciato}(2012){Genel}, {Dekel}, \&
  {Cacciato}}]{gdc12}
{Genel} S., {Dekel} A., {Cacciato} M., 2012, \mnras, 425, 788

\bibitem[{{Genel} {et~al}\mbox{.}(2012){Genel}, {Naab}, {Genzel}, {F{\"o}rster
  Schreiber}, {Sternberg}, {Oser}, {Johansson}, {Dav{\'e}}, {Oppenheimer}, \&
  {Burkert}}]{genel12}
{Genel} S. {et~al.}, 2012, \apj, 745, 11

\bibitem[{{Genzel} {et~al}\mbox{.}(2008){Genzel}, {Burkert}, {Bouch{\'e}},
  {Cresci}, {F{\"o}rster Schreiber}, {Shapley}, {Shapiro}, {Tacconi}, \& {et
  al.,}}]{genzel08}
{Genzel} R. {et~al.}, 2008, \apj, 687, 59

\bibitem[{{Genzel} {et~al}\mbox{.}(2013){Genzel}, {F{\"o}rster Schreiber},
  {Lang}, {Tacchella}, {Tacconi}, {Wuyts}, \& {et al.}}]{genzel13}
{Genzel} R., {F{\"o}rster Schreiber} N.~M., {Lang} P., {Tacchella} S.,
  {Tacconi} L.~J., {Wuyts} S., {et al.}, 2013, arXiv:1310.3838

\bibitem[{{Genzel} {et~al}\mbox{.}(2011){Genzel}, {Newman}, {Jones},
  {F{\"o}rster Schreiber}, {Shapiro}, {Genel}, {Lilly}, \& {et al.}}]{genzel11}
{Genzel} R., {Newman} S., {Jones} T., {F{\"o}rster Schreiber} N.~M., {Shapiro}
  K., {Genel} S., {Lilly} S.~J., {et al.}, 2011, \apj, 733, 101

\bibitem[{{Genzel} {et~al}\mbox{.}(2006){Genzel}, {Tacconi}, {Eisenhauer},
  {F{\"o}rster Schreiber}, {Cimatti}, {Daddi}, {Bouch{\'e}}, \& {et
  al.}}]{genzel06}
{Genzel} R., {Tacconi} L.~J., {Eisenhauer} F., {F{\"o}rster Schreiber} N.~M.,
  {Cimatti} A., {Daddi} E., {Bouch{\'e}} N., {et al.}, 2006, \nat, 442, 786

\bibitem[{{Hopkins} {et~al}\mbox{.}(2006){Hopkins}, {Hernquist}, {Cox},
  {Robertson}, \& {Springel}}]{hopkins06}
{Hopkins} P.~F., {Hernquist} L., {Cox} T.~J., {Robertson} B., {Springel} V.,
  2006, \apjs, 163, 50

\bibitem[{{Hopkins} {et~al}\mbox{.}(2012){Hopkins}, {Kere{\v s}}, {Murray},
  {Quataert}, \& {Hernquist}}]{hopkins12c}
{Hopkins} P.~F., {Kere{\v s}} D., {Murray} N., {Quataert} E., {Hernquist} L.,
  2012, \mnras, 427, 968

\bibitem[{{Hopkins}, {Quataert} \& {Murray}(2012){Hopkins}, {Quataert}, \&
  {Murray}}]{hopkins12b}
{Hopkins} P.~F., {Quataert} E., {Murray} N., 2012, \mnras, 421, 3522

\bibitem[{{Immeli} {et~al}\mbox{.}(2004{\natexlab{a}}){Immeli}, {Samland},
  {Gerhard}, \& {Westera}}]{immeli04_b}
{Immeli} A., {Samland} M., {Gerhard} O., {Westera} P., 2004{\natexlab{a}},
  \aap, 413, 547

\bibitem[{{Immeli} {et~al}\mbox{.}(2004{\natexlab{b}}){Immeli}, {Samland},
  {Westera}, \& {Gerhard}}]{immeli04_a}
{Immeli} A., {Samland} M., {Westera} P., {Gerhard} O., 2004{\natexlab{b}},
  \apj, 611, 20

\bibitem[{{Kannappan}(2004)}]{kannappan04}
{Kannappan} S.~J., 2004, \apjl, 611, L89

\bibitem[{{Kauffmann} {et~al}\mbox{.}(2012){Kauffmann}, {Li}, {Fu},
  {Saintonge}, {Catinella}, {Tacconi}, {Kramer}, \& {et al.}}]{kauffmann12}
{Kauffmann} G., {Li} C., {Fu} J., {Saintonge} A., {Catinella} B., {Tacconi}
  L.~J., {Kramer} C., {et al.}, 2012, \mnras, 422, 997

\bibitem[{{Kaviraj} {et~al}\mbox{.}(2013{\natexlab{a}}){Kaviraj}, {Cohen},
  {Ellis}, {Peirani}, {Windhorst}, {O'Connell}, {Silk}, {Whitmore}, {Hathi},
  {Ryan}, {Dopita}, {Frogel}, \& {Dekel}}]{kaviraj13a}
{Kaviraj} S. {et~al.}, 2013{\natexlab{a}}, \mnras, 428, 925

\bibitem[{{Kaviraj} {et~al}\mbox{.}(2013{\natexlab{b}}){Kaviraj}, {Cohen},
  {Windhorst}, {Silk}, {O'Connell}, {Dopita}, {Dekel}, {Hathi}, {Straughn}, \&
  {Rutkowski}}]{kaviraj13b}
{Kaviraj} S. {et~al.}, 2013{\natexlab{b}}, \mnras, 429, L40

\bibitem[{{Kere{\v s}} {et~al}\mbox{.}(2009){Kere{\v s}}, {Katz}, {Fardal},
  {Dav{\'e}}, \& {Weinberg}}]{keres09}
{Kere{\v s}} D., {Katz} N., {Fardal} M., {Dav{\'e}} R., {Weinberg} D.~H., 2009,
  \mnras, 395, 160

\bibitem[{{Kere{\v s}} {et~al}\mbox{.}(2005){Kere{\v s}}, {Katz}, {Weinberg},
  \& {Dav{\'e}}}]{keres05}
{Kere{\v s}} D., {Katz} N., {Weinberg} D.~H., {Dav{\'e}} R., 2005, \mnras, 363,
  2

\bibitem[{{Khochfar} \& {Ostriker}(2008)}]{khochfar08}
{Khochfar} S., {Ostriker} J.~P., 2008, \apj, 680, 54

\bibitem[{{Kimm} {et~al}\mbox{.}(2011){Kimm}, {Devriendt}, {Slyz}, {Pichon},
  {Kassin}, \& {Dubois}}]{kimm11}
{Kimm} T., {Devriendt} J., {Slyz} A., {Pichon} C., {Kassin} S.~A., {Dubois} Y.,
  2011, arXiv:1106.0538

\bibitem[{{Krumholz} \& {Burkert}(2010)}]{krum_burkert10}
{Krumholz} M.~R., {Burkert} A., 2010, \apj, 724, 895

\bibitem[{{Krumholz} \& {Dekel}(2010)}]{kd10}
{Krumholz} M.~R., {Dekel} A., 2010, \mnras, 406, 112

\bibitem[{{Krumholz} \& {Dekel}(2012)}]{kd12}
{Krumholz} M.~R., {Dekel} A., 2012, \apj, 753, 16

\bibitem[{{Krumholz}, {Dekel} \& {McKee}(2012){Krumholz}, {Dekel}, \&
  {McKee}}]{kdm12}
{Krumholz} M.~R., {Dekel} A., {McKee} C.~F., 2012, \apj, 745, 69

\bibitem[{{Krumholz} \& {Thompson}(2013)}]{krum_thom13}
{Krumholz} M.~R., {Thompson} T.~A., 2013, \mnras, 434, 2329

\bibitem[{{Larson}(1976)}]{larson76}
{Larson} R.~B., 1976, \mnras, 176, 31

\bibitem[{{Law} {et~al}\mbox{.}(2012){Law}, {Steidel}, {Shapley}, {Nagy},
  {Reddy}, \& {Erb}}]{law12}
{Law} D.~R., {Steidel} C.~C., {Shapley} A.~E., {Nagy} S.~R., {Reddy} N.~A.,
  {Erb} D.~K., 2012, \apj, 759, 29

\bibitem[{{Lee} {et~al}\mbox{.}(2013){Lee}, {Giavalisco}, {Williams}, {Guo},
  {Lotz}, {Van der Wel}, {Ferguson}, \& {et al.}}]{lee13}
{Lee} B., {Giavalisco} M., {Williams} C.~C., {Guo} Y., {Lotz} J., {Van der Wel}
  A., {Ferguson} H.~C., {et al.}, 2013, \apj, 774, 47

\bibitem[{{Mac Low}(1999)}]{maclow99}
{Mac Low} M.-M., 1999, \apj, 524, 169

\bibitem[{{Maller} \& {Dekel}(2002)}]{maller02}
{Maller} A.~H., {Dekel} A., 2002, \mnras, 335, 487

\bibitem[{{Martig} {et~al}\mbox{.}(2009){Martig}, {Bournaud}, {Teyssier}, \&
  {Dekel}}]{martig09}
{Martig} M., {Bournaud} F., {Teyssier} R., {Dekel} A., 2009, \apj, 707, 250

\bibitem[{{Martig} {et~al}\mbox{.}(2013){Martig}, {Crocker}, {Bournaud},
  {Emsellem}, {Gabor}, {Alatalo}, {Blitz}, \& {et al.}}]{martig13}
{Martig} M., {Crocker} A.~F., {Bournaud} F., {Emsellem} E., {Gabor} J.~M.,
  {Alatalo} K., {Blitz} L., {et al.}, 2013, \mnras, 432, 1914

\bibitem[{{Mihos} \& {Hernquist}(1996)}]{mihos96}
{Mihos} J.~C., {Hernquist} L., 1996, \apj, 464, 641

\bibitem[{{Mo}, {Mao} \& {White}(1998){Mo}, {Mao}, \& {White}}]{mmw98}
{Mo} H.~J., {Mao} S., {White} S.~D.~M., 1998, \mnras, 295, 319

\bibitem[{{Murray}, {Quataert} \& {Thompson}(2005){Murray}, {Quataert}, \&
  {Thompson}}]{murray05}
{Murray} N., {Quataert} E., {Thompson} T.~A., 2005, \apj, 618, 569

\bibitem[{{Murray}, {Quataert} \& {Thompson}(2010){Murray}, {Quataert}, \&
  {Thompson}}]{murray10}
{Murray} N., {Quataert} E., {Thompson} T.~A., 2010, \apj, 709, 191

\bibitem[{{Neistein} \& {Dekel}(2008)}]{neistein08b}
{Neistein} E., {Dekel} A., 2008, \mnras, 388, 1792

\bibitem[{{Neistein}, {van den Bosch} \& {Dekel}(2006){Neistein}, {van den
  Bosch}, \& {Dekel}}]{neistein06}
{Neistein} E., {van den Bosch} F.~C., {Dekel} A., 2006, \mnras, 372, 933

\bibitem[{{Nelson} {et~al}\mbox{.}(2013){Nelson}, {Vogelsberger}, {Genel},
  {Sijacki}, {Kere{\v s}}, {Springel}, \& {Hernquist}}]{nelson13}
{Nelson} D., {Vogelsberger} M., {Genel} S., {Sijacki} D., {Kere{\v s}} D.,
  {Springel} V., {Hernquist} L., 2013, \mnras, 429, 3353

\bibitem[{{Newman} {et~al}\mbox{.}(2010){Newman}, {Ellis}, {Treu}, \&
  {Bundy}}]{newman10}
{Newman} A.~B., {Ellis} R.~S., {Treu} T., {Bundy} K., 2010, \apjl, 717, L103

\bibitem[{{Newman} {et~al}\mbox{.}(2013){Newman}, {Genzel}, {F{\"o}rster
  Schreiber}, {Shapiro Griffin}, {Mancini}, {Lilly}, {Renzini}, {Bouch{\'e}},
  \& {et al.}}]{newman13}
{Newman} S.~F. {et~al.}, 2013, \apj, 767, 104

\bibitem[{{Noguchi}(1999)}]{noguchi99}
{Noguchi} M., 1999, \apj, 514, 77

\bibitem[{{Ocvirk}, {Pichon} \& {Teyssier}(2008){Ocvirk}, {Pichon}, \&
  {Teyssier}}]{ocvirk08}
{Ocvirk} P., {Pichon} C., {Teyssier} R., 2008, \mnras, 390, 1326

\bibitem[{{Pichon} {et~al}\mbox{.}(2011){Pichon}, {Pogosyan}, {Kimm}, {Slyz},
  {Devriendt}, \& {Dubois}}]{pichon11}
{Pichon} C., {Pogosyan} D., {Kimm} T., {Slyz} A., {Devriendt} J., {Dubois} Y.,
  2011, \mnras, 418, 2493

\bibitem[{{Saintonge} {et~al}\mbox{.}(2012){Saintonge}, {Tacconi}, {Fabello},
  {Wang}, {Catinella}, {Genzel}, {Graci{\'a}-Carpio}, \& {et
  al.}}]{saintonge12}
{Saintonge} A., {Tacconi} L.~J., {Fabello} S., {Wang} J., {Catinella} B.,
  {Genzel} R., {Graci{\'a}-Carpio} J., {et al.}, 2012, \apj, 758, 73

\bibitem[{{Stewart} {et~al}\mbox{.}(2013){Stewart}, {Brooks}, {Bullock},
  {Maller}, {Diemand}, {Wadsley}, \& {Moustakas}}]{stewart13}
{Stewart} K.~R., {Brooks} A.~M., {Bullock} J.~S., {Maller} A.~H., {Diemand} J.,
  {Wadsley} J., {Moustakas} L.~A., 2013, \apj, 769, 74

\bibitem[{{Tacconi} {et~al}\mbox{.}(2010){Tacconi}, {Genzel}, {Neri}, {Cox},
  {Cooper}, {Shapiro}, {Bolatto}, {Bouch{\'e}}, \& {et al.}}]{tacconi10}
{Tacconi} L.~J. {et~al.}, 2010, \nat, 463, 781

\bibitem[{{Tacconi} {et~al}\mbox{.}(2013){Tacconi}, {Neri}, {Genzel}, {Combes},
  {Bolatto}, {Cooper}, {Wuyts}, \& {et al.}}]{tacconi13}
{Tacconi} L.~J., {Neri} R., {Genzel} R., {Combes} F., {Bolatto} A., {Cooper}
  M.~C., {Wuyts} S., {et al.}, 2013, \apj, 768, 74

\bibitem[{{Tillson} {et~al}\mbox{.}(2012){Tillson}, {Devriendt}, {Slyz},
  {Miller}, \& {Pichon}}]{tillson12}
{Tillson} H., {Devriendt} J., {Slyz} A., {Miller} L., {Pichon} C., 2012,
  arXiv:1211.3124

\bibitem[{{Toomre}(1964)}]{toomre64}
{Toomre} A., 1964, \apj, 139, 1217

\bibitem[{{van der Wel} {et~al}\mbox{.}(2011){van der Wel}, {Rix}, {Wuyts},
  {McGrath}, {Koekemoer}, {Bell}, {Holden}, {Robaina}, \&
  {McIntosh}}]{vanderwel11}
{van der Wel} A. {et~al.}, 2011, \apj, 730, 38

\bibitem[{{van Dokkum} {et~al}\mbox{.}(2008){van Dokkum}, {Franx}, {Kriek},
  {Holden}, {Illingworth}, {Magee}, {Bouwens}, {Marchesini}, {Quadri},
  {Rudnick}, {Taylor}, \& {Toft}}]{dokkum08}
{van Dokkum} P.~G. {et~al.}, 2008, \apjl, 677, L5

\bibitem[{{van Dokkum}, {Kriek} \& {Franx}(2009){van Dokkum}, {Kriek}, \&
  {Franx}}]{dokkum09}
{van Dokkum} P.~G., {Kriek} M., {Franx} M., 2009, \nat, 460, 717

\bibitem[{{Whitaker} {et~al}\mbox{.}(2012){Whitaker}, {Kriek}, {van Dokkum},
  {Bezanson}, {Brammer}, {Franx}, \& {Labb{\'e}}}]{whitaker12}
{Whitaker} K.~E., {Kriek} M., {van Dokkum} P.~G., {Bezanson} R., {Brammer} G.,
  {Franx} M., {Labb{\'e}} I., 2012, \apj, 745, 179

\bibitem[{{Williams} {et~al}\mbox{.}(2013){Williams}, {Giavalisco}, {Cassata},
  {Tundo}, {Wiklind}, {Guo}, {Lee}, ...~{Dekel}, \& {et al.,}}]{williams13}
{Williams} C.~C. {et~al.}, 2013, arXiv:1310.3819

\end{thebibliography}


\label{lastpage}
\end{document}